\documentclass[epj]{webofc}
\usepackage[varg]{txfonts}   
\wocname{EPJ Web of Conferences}
\woctitle{CONF12}
\begin{document}
\selectlanguage{english}
\title{Quark Matter Equation of State from Perturbative QCD}
%
%

\author{Aleksi Vuorinen \inst{1}\fnsep\thanks{\email{aleksi.vuorinen@helsinki.fi}}
}

\institute{Dept.~of Physics and Helsinki Institute of Physics, P.O.~Box 64, FI-00014 University of Helsinki, Finland
}

\abstract{%
  In this proceedings contribution, we discuss recent developments in the perturbative determination of the Equation of State of dense quark matter, relevant for the microscopic description of neutron star cores. First, we introduce the current state of the art in the problem, both at zero and small temperatures, and then present results from two recent perturbative studies that pave the way towards extending the EoS to higher orders in perturbation theory.
}
\maketitle
\section{Introduction}
\label{intro}

The cores of neutron stars contain some of the densest matter in our Universe, second only to the interiors of black holes \cite{Lattimer:2004pg,Brambilla:2014jmp}. Predicting the collective properties of this extreme form of matter is a well-known problem in the realm of nuclear physics and QCD, but unfortunately still lacks a satisfactory solution due to the absence of nonperturbative first principles methods applicable to it. With lattice QCD suffering from the infamous Sign Problem \cite{deForcrand:2010ys} and the applicability of Chiral Effective Theory (CET) extending only up to roughly the nuclear matter saturation density \cite{Tews:2012fj}, there is an urgent need to find complementary methods for tackling strongly coupled matter at the the highest densities reached within the stars. This has motivated attempts to approach the problem using methods ranging from the Functional Renormalization Group to phenomenological models and the holographic duality \cite{Drews:2016wpi,Buballa:2003qv,Li:2015uea,Preis:2016fsp,Hoyos:2016cob,Hoyos:2016zke}, but they all come with their own systematic uncertainties and limitations.

In these conference proceedings, we describe recent efforts to approach the problem of quark matter using insights gained from the limit of very high densities, where a weak coupling approach is guaranteed to be valid due to the asymptotic freedom of the underlying theory. It has been demonstrated in a simple setup that such high-density information can be efficiently used to constrain the behavior of strongly interacting matter at densities realized inside neutron star cores \cite{Kurkela:2014vha}, and work is currently underway to combine these insights with state-of-the-art Bayesian analyses taking into account all available input from astrophysical observations to low-density calculations \cite{annalanattila}. To aid this process, it is clearly imperative to vigorously work on improving the current state-of-the-art results for the perturbative Equation of State (EoS) of dense quark matter.

This article is structured as follows. In section 2, we introduce the current state of the art for the perturbative EoS of quark matter and briefly discuss applications of these results to neutron star physics. In section 3, we then dwell into some recently completed work that generalized existing zero-temperature EoS calculations to small but nonzero $T$, while in section 4 lay out a roadmap for extending the $T=0$ EoS to the full next order in a weak coupling expansion. Brief concluding remarks are finally made in section 5.

\section{State of the art}

\begin{figure}[t]
\centering
\includegraphics[width=11cm,clip]{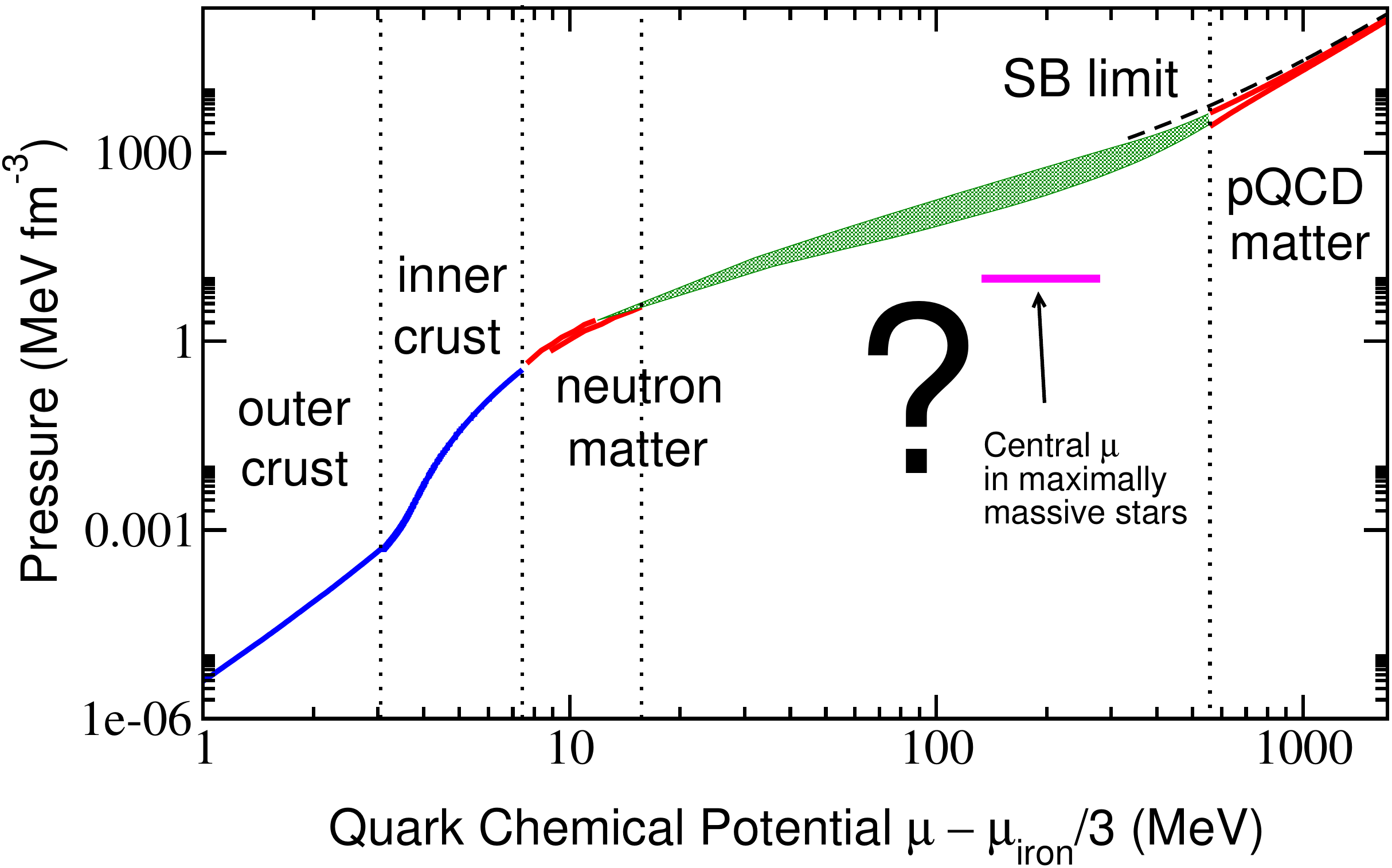}
\caption{A cartoon depicting the behavior of the neutron star matter EoS as a function of the (shifted) quark chemical potential  $\mu$. The blue curve on the left is taken from \cite{Negele:1971vb,Ruester:2005fm}, the two red curves on the left from \cite{Tews:2012fj}, the red curves on the right from \cite{Kurkela:2009gj}, and the green band from an interpolation performed in \cite{Kurkela:2014vha}. The figure itself is taken from the last of these references.}
\label{fig-1}       
\end{figure}

In figure 1, we depict the current state of the art in our understanding of the EoS of cold and dense QCD matter as relevant for neutron stars. Starting from the lowest densities on the left side of the figure, the EoS of the outer and inner crust regions is accurately described using well-established methods of nuclear physics, such as strongly constrained potential models \cite{Negele:1971vb,Ruester:2005fm}. Difficulties emerge, however, when we approach the outer core of the star, corresponding to densities close to the nuclear matter saturation density $n_s$. Here, the CET framework has turned out to be a highly valuable tool, providing both access to densities somewhat beyond $n_s$ and a first-principles understanding of the hierarchy between various types of operators composed of the nucleonic degrees of freedom in the system (see e.g.~\cite{Tews:2012fj} and references therein). However, even CET has its limitations, and proceeding much beyond the saturation density with these medhods appears extremely nontrivial, although some promising progress has been achieved very recently \cite{Drischler:2016djf}.

Starting from the opposite end of extremely high densities, the asymptotic freedom of QCD guarantees that a perturbative QCD (pQCD) approach to the problem be feasible. Unfortunately, the sizable value of the strong coupling constant implies that one needs to go to a relatively high order in the weak coupling expansion of the EoS in order to extend the applicability of the perturbative result to a region of relevance for neutron star physics. This is well demonstrated in fig.~2, where the current perturbative EoS of $T=0$ quark matter is depicted  as a function of baryon chemical potential. Unlike in the case of high temperatures where the well-developed tools of resummed perturbative theory have enabled a successful description of the EoS down to temperatures close to the (pseudo)critical temperature of the deconfinement transition (left) \cite{klry,Andersen:2011sf}, the situation at zero temperature and high density is still less than optimal (right). Here, the current state-of-the-art result is from a three-loop computation including the effects of finite quark masses \cite{Kurkela:2009gj}, which builds on a series of earlier works  \cite{fmcl,avpres} and is visible as the rightmost red band in fig.~1.

\begin{figure}[t]
\centering
\includegraphics[width=6.5cm,clip]{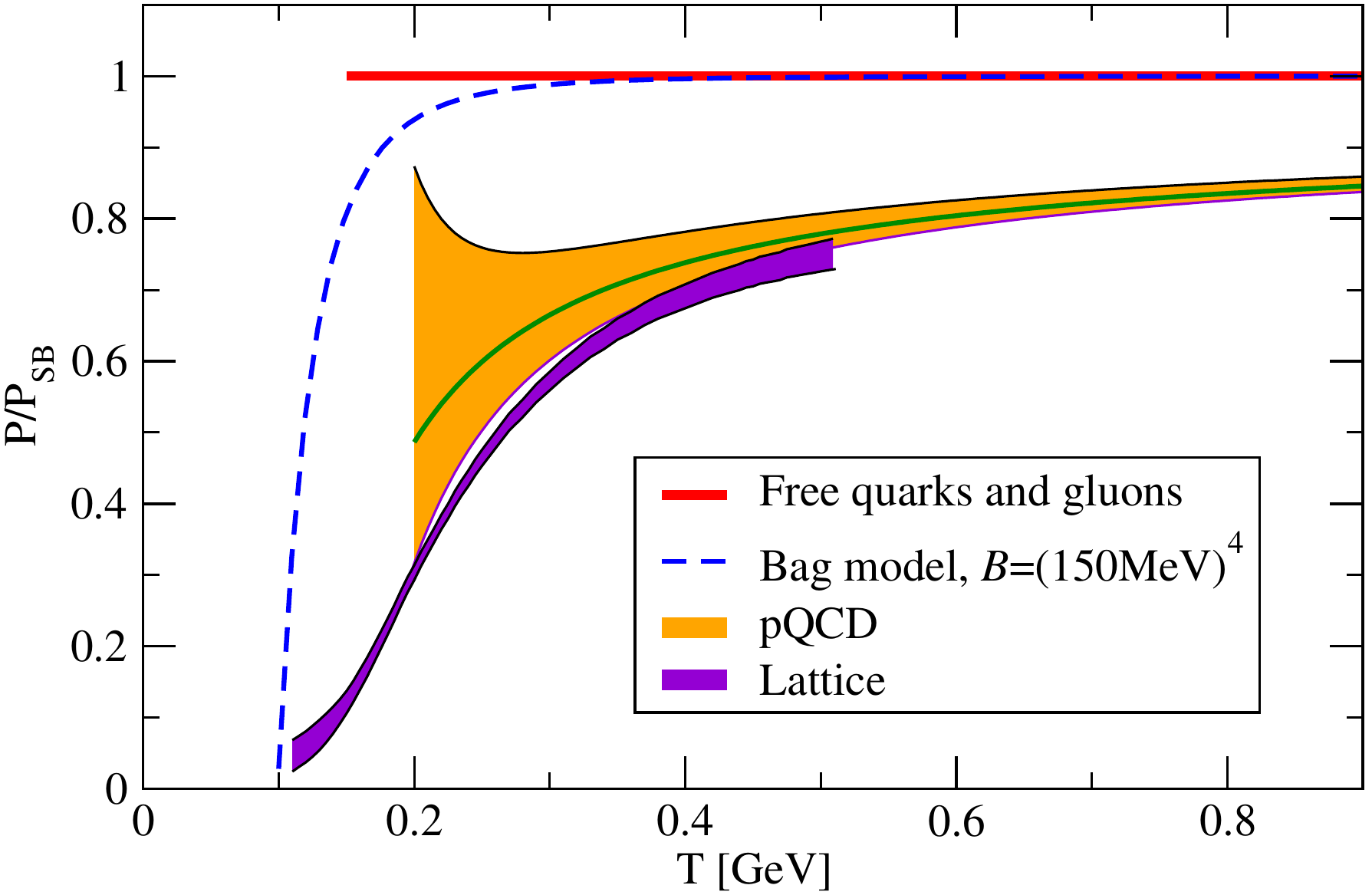}$\;\;$
\includegraphics[width=6.5cm,clip]{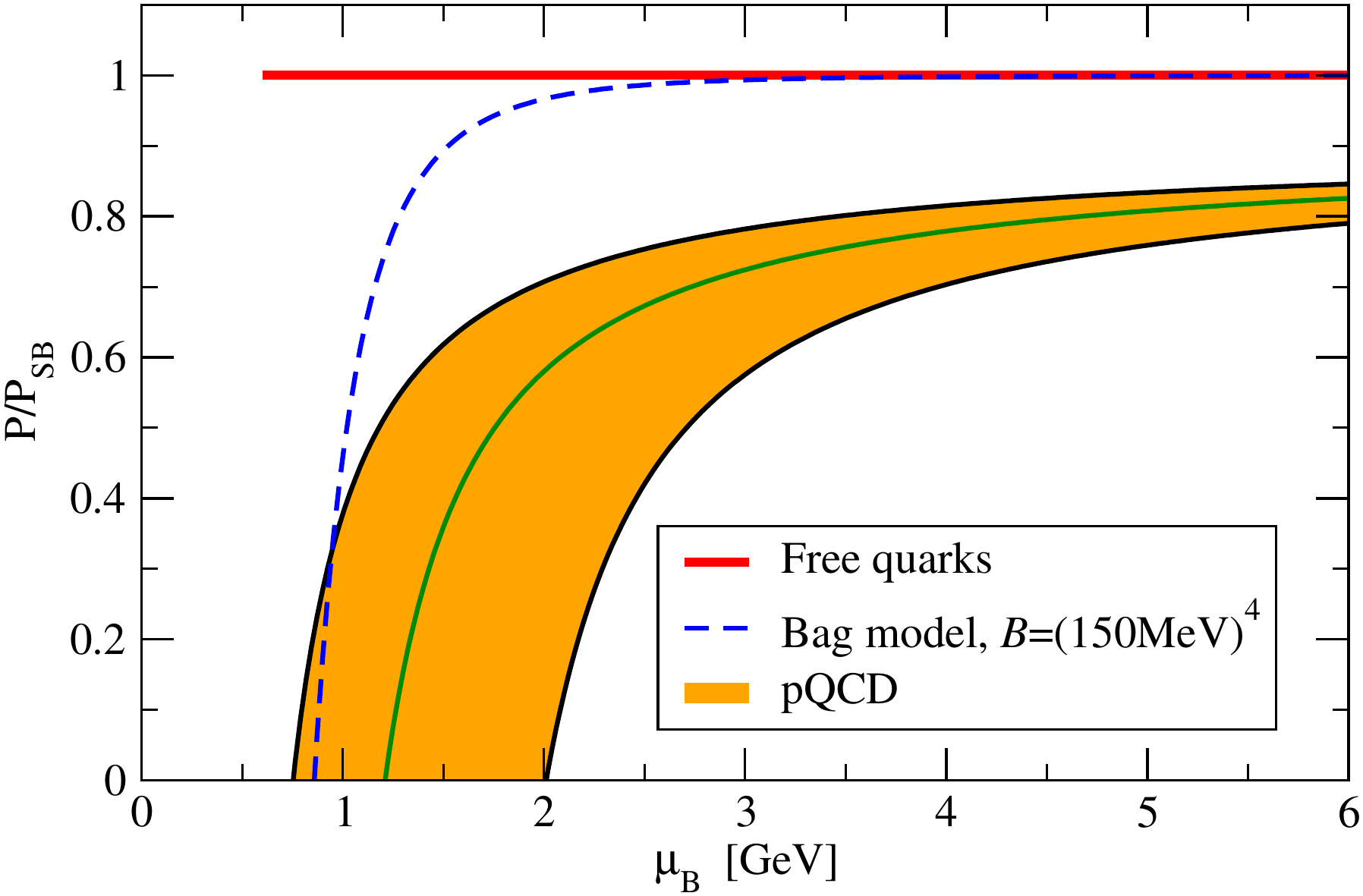}
\caption{Left: the EoS of $\mu=0$ quark gluon plasma as a function of temperature, with state-of-the-art lattice  \cite{Borsanyi:2010cj, Borsanyi:2013bia} and pQCD \cite{klry} results plotted together with the prediction of the MIT bag model. Right: The zero-temperature EoS of quark matter, obtained from the three-loop computation of \cite{Kurkela:2009gj}. In both cases, the perturbative bands are obtained by varying the renormalization scale by a factor of two around a midpoint value. Both plots are taken from \cite{Fraga:2013qra}.}
\label{fig-2}       
\end{figure}

Despite the difficulties in both the CET and pQCD approaches, it has been shown that taking the state-of-the-art results from both methods, it is possible to efficiently constrain the behavior of the neutron star matter EoS at all densities. An interpolating EoS, utilizing the results of \cite{Tews:2012fj,Kurkela:2009gj} and constructed demanding only thermodynamic consistency, subluminality and the ability to support a two-solar-mass star, is shown as the green band in fig.~1. This prediction has an uncertainty of only ca.~$\pm 40\%$ at worst, which is quite a remarkable result. In particular, the fact that the high-density constraint is important even at densities realizable inside real life neutron stars is far from obvious, as discussed at some length in \cite{Kurkela:2014vha} (cf.~also the related work of \cite{Hebeler:2013nza}).

In order to gain a quantitative understanding of the neutron star matter EoS using only robust first principles tools, vigorous work must clearly be performed to shrink the gap between the low- and high-density results depicted in fig.~1. In the high-density regime, this implies extending the current three-loop EoS of cold quark matter to the next perturbative order, i.e.~to the full four-loop accuracy. This is, however, a formidable task, and necessitates both the development of novel tools for multi-loop diagrammatic calculations at finite density and gaining a better understanding of the soft sector of QCD at high density. First steps in this direction have nevertheless already been taken, and in the following two sections we review two recent advances in the field.

\section{Cool quark matter}

While old, quiescent neutron stars have cooled down enough to allow neglecting thermal effects, the description of supernova explosions as well mergers of two neutron stars necessitate maintaining a nonzero $T$ in the determination of the EoS \cite{Shen:1998gq}. With the highest temperatures reached in mergers estimated around 100 MeV, it becomes clear that the behavior of the system can in fact be radically different from its Fermi sphere dominated counterpart at zero temperature. This makes it essential to also revisit the perturbative determination of the EoS of dense quark matter, and try to merge the gap between the $T=0$ result and the regime of high temperatures.

Due to technical difficulties associated with the description of the IR sector of QCD at small but nonzero temperatures, this region of the phase diagram was largely neglected for a long time. In the high-temperature regime, characterized by $T\geq g\mu$, the weak coupling expansion of the EoS has been worked out up to partial four-loop order, or ${\mathcal O}(g^6\ln\,g)$ in the gauge coupling \cite{avpres}. At the same time, only an ${\mathcal O}(g^3)$ calculation connected this result to the $T=0$ limit until relatively recently \cite{Ipp:2006ij}. Even then, the situation was first remedied in a rather cumbersome way by performing an explicit resummation of an infinite class of Feynman diagrams in full QCD, which lead to an ${\mathcal O}(g^4)$ result applicable throughout the deconfined phase of the theory \cite{Ipp:2006ij}.

\begin{figure}[t]
\centering
\includegraphics[width=13cm,clip]{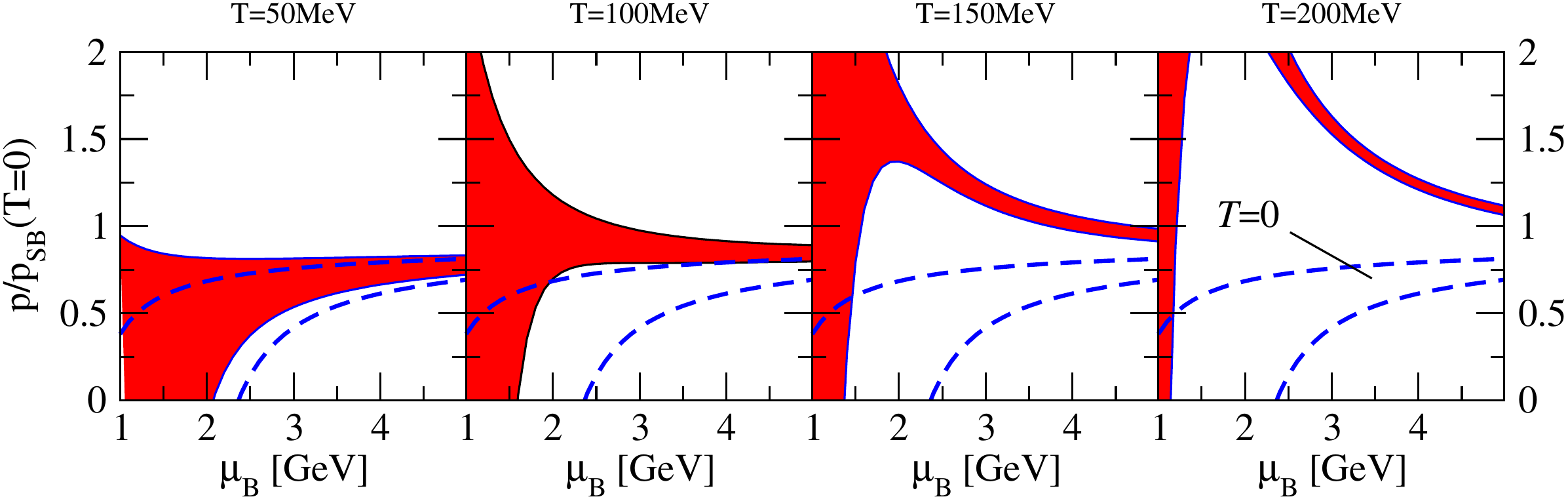}
\caption{The pressure of QCD at four fixed nonzero values of the temperature. The red band is again obtained upon varying the scale parameter, and the dashed blue line shows the result at $T=0$. The figure is from \cite{Kurkela:2016was}.}
\label{fig-3}       
\end{figure}

In a recent article \cite{Kurkela:2016was}, we revisited the problem of determining the EoS of deconfined QCD matter at all temperatures, only this time from a somewhat different standpoint. We added and subtracted  to/from the QCD pressure a function that by definition represents the contributions of the IR sensitive field modes that need to be resummed. This lead to the seemingly trivial identity, resembling a trick used e.g.~in \cite{Laine:2011xm},
\begin{eqnarray}
p_\text{QCD}^\text{res} \;= \;p_\text{QCD}^\text{res} - p_\text{soft}^\text{res} + p_\text{soft}^\text{res} 
\;=\;  p_\text{QCD}^\text{naive} - p_\text{soft}^\text{naive} + p_\text{soft}^\text{res}\,,  \label{starting}
\end{eqnarray}
where `res' refers to resummed expressions and `naive' to ones evaluated in a strict loop expansion. Despite its benign appearance, the step taken at the second equal sign is highly important: we have noted here that the difference between the quantities $p_\text{QCD}^\text{res}$ and $p_\text{soft}^\text{res}$ is by definition IR safe, implying that both functions can be computed in a naive weak coupling expansion. This is a tremendous simplification, as it means that to order $g^4$, the pressure of the full theory, $p_\text{QCD}$, can be directly taken over from the literature \cite{avpres} and we only need to worry about the proper identification of the soft sector of the theory and the correct determination of its contribution to the pressure.

The first question to address becomes now, what degrees of freedom constitute the soft sector of QCD at different values of $T$ and $\mu$, and what is the effective theory that ought to be used to describe them. It turns out that a choice sufficient for obtaining the pressure to order $g^5$ at all temperatures and densities is to apply the dimensionally reduced (DR) effective theory EQCD to the description of the static bosonic sector of the theory \cite{Appelquist:1981vg,Kajantie:1995dw,Braaten:1995cm}, and to treat the soft non-static modes with $k\sim g$ using the  HTL effective theory \cite{Braaten:1989mz,Braaten:1991gm}. This leads to the result
\begin{eqnarray}
p_\text{QCD} &=& p_\text{QCD}^\text{naive} + p_\text{DR}^\text{res} - p_\text{DR}^\text{naive} + p_\text{HTL}^\text{res} -  p_\text{HTL}^\text{naive}\, , \label{resgen}
\end{eqnarray}
where it should be noted that the HTL resummation is only applied to the non-static modes. The physical nature and properties of each of these terms is discussed in quite some detail in \cite{Kurkela:2016was}. Here, we merely note that
\begin{itemize}
\item With the exception of the very last term, $p_\text{HTL}^\text{naive}$, each of the pieces in the above result can be obtained directly from the literature, cf.~e.g.~\cite{avpres,Andersen:1999sf}.
\item All UV and IR divergences cancel between the above terms upon renormalization. The exact cancellation pattern is somewhat nontrivial, featuring both $1/\epsilon$ terms, generated by dimensional regularization, and logarithms of the temperature that diverge in the $T\to 0$ limit. 
\end{itemize}

In figs.~3 and 4, we display the behavior of the above results in two different ways. In fig.~3, resembling fig.~2.b, we show the behavior of the pressure as a function of the baryon chemical potential for four fixed nonzero values of the temperature. Note that the pressure is normalized here to that of the non-interacting theory at $T=0$, while the interacting $T=0$ result is shown as the dashed blue curves in the figures. To demonstrate the successful interpolation of our result between known limits, in fig.~4 we next display the behavior of the pressure as a function of temperature at two fixed values of $T^2+(\mu_B/3\pi)^2$. We observe that at extremely small temperatures the behavior of the result can be captured by the Hard Dense Loop results of \cite{Gerhold:2004tb,Gerhold:2005uu}, while at larger values of $T$ it quickly approaches that of the high-$T$ computation of \cite{avpres}.

\begin{figure}[t]
\centering
\includegraphics[width=6.5cm,clip]{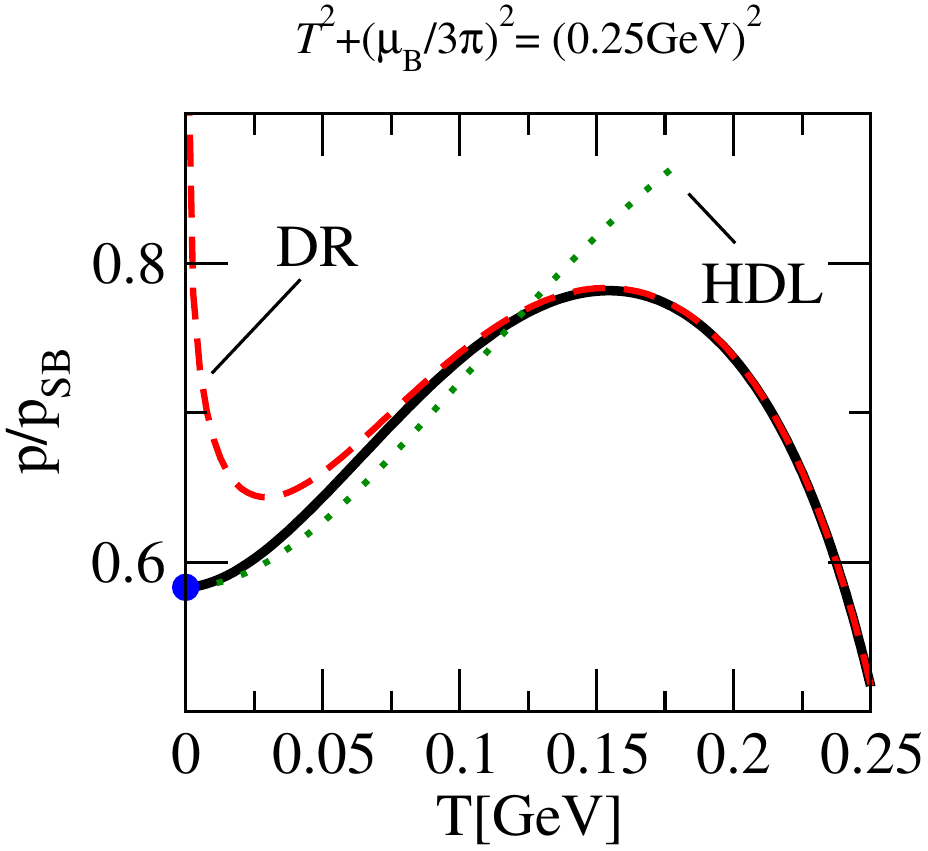}$\;\;$
\includegraphics[width=6.5cm,clip]{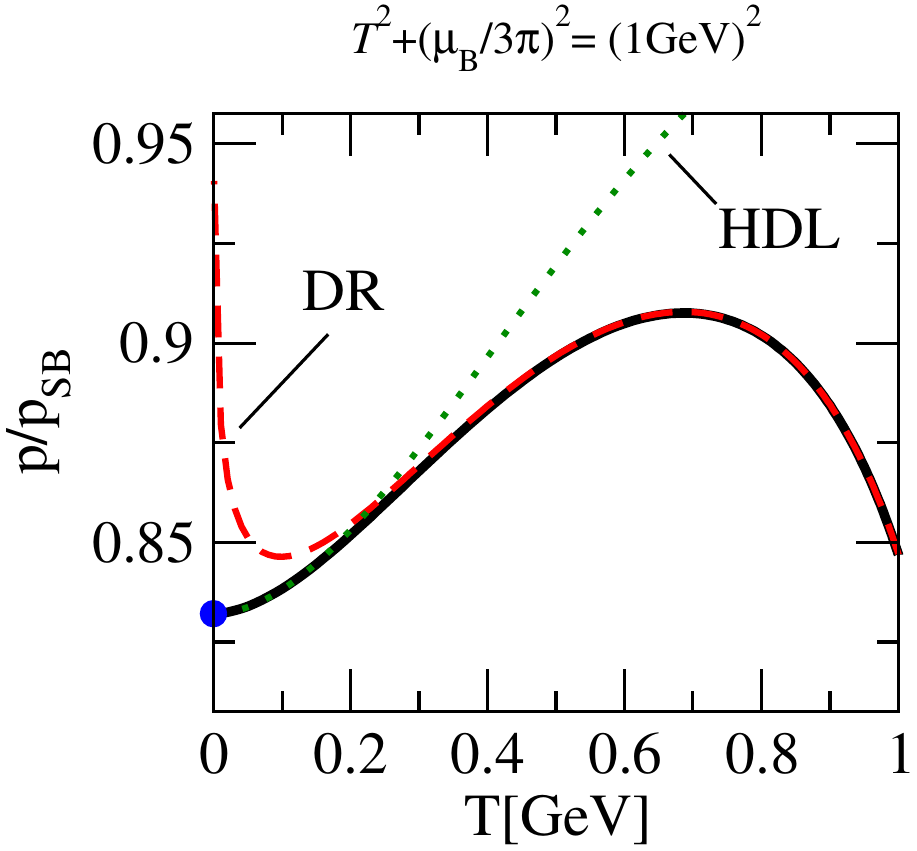}
\caption{The pressure of QCD as computed in \cite{Kurkela:2016was}, compared with the known low- and high-$T$ limits. See the main text for a more detailed explanation. The figure is from \cite{Kurkela:2016was}.}
\label{fig-4}       
\end{figure}

\section{Towards the four-loop EoS at zero temperature}

Returning now back to systems, where thermal effects can be altogether neglected, we are faced with the challenges laid out in sec.~2. In this context, an extremely challenging, yet realistic goal is to attempt extending the EoS of $T=0$ quark matter to the full four-loop order in perturbation theory, including terms of orders $g^6\ln^2\,g$, $g^6\ln\,g$, and $g^6$ in the coupling constant. This necessitates solving two somewhat separate problems: first, how to organize the calculation in such a way that the contributions of all momentum scales are properly accounted for, and second, how to efficiently perform high loop order computations at zero temperature and finite density. In the two subsections below, we shall briefly address both of these issues.

\subsection{Outline of the computation and soft contributions}

Our starting point in the evaluation of the four-loop EoS of zero-temperature quark matter is again eq.~(2), but recalling now that the DR contribution vanishes at $T=0$, producing the expression
\begin{eqnarray}
p_\text{QCD} &=& p_\text{QCD}^\text{naive} + p_\text{HTL}^\text{res} -  p_\text{HTL}^\text{naive}\,. \label{res2}
\end{eqnarray}
While the first term here is conceptually simple --- obtainable by evaluating all vacuum diagrams in the theory up to the desired loop order --- the HTL parts require a more careful consideration.

The Hard Thermal Loop effective theory is known to correctly capture the dynamics of the soft scales of deconfined QCD at zero temperature and finite density, and thus a successful evaluation of the partition function of this theory is sufficient for obtaining the soft contributions to the QCD pressure. The beauty of eq.~(\ref{res2}) above is that the HTL contribution enters in the form of the \textit{difference} of its resummed and naive forms, which guarantees that no double counting of contributions can take place. Should some degrees of freedom in the effective theory not require resummation, their contribution to the pressure becomes canceled in this difference. This fact may in fact be used to simplify the evaluation of $p_\text{HTL}$, and it can e.g.~be shown that it is not necessary to consider high-order fermion self energies or vertex functions in the HTL theory as long as one is only interested in the pressure up to the $g^6$ order. 

Of the two terms to be computed, it suffices to consider the resummed version, as its naive counterpart can be obtained from this expression in a straightforward manner. As we shall show in detail in a forthcoming publication \cite{newpaper}, the most nontrivial part of the calculation boils down to determining the two-loop HTL gluon polarization tensor and the Next-to-Leading Order asymptotic mass parameter of the gluon field. This is, however, a substantial challenge that in particular requires carrying out a separate computation in resummed perturbation theory. The details of this exercise are left to be thoroughly explained in our later publication \cite{newpaper}.

\subsection{Hard contributions}

Moving on to the first term of eq.~(\ref{res2}), $p_\text{QCD}^\text{naive}$, we are faced with the challenge of evaluating all four-loop vacuum diagrams of QCD at zero temperature and finite chemical potentials. This is a formidable challenge and clearly requires an automated approach to become feasible --- despite the fact that a number of digrams can be discarded due to containing no fermion loops and thus no $\mu$ dependence. A crucial aid in this process turns out to be an analytic tool dubbed `cutting rules'. These rules were first discussed in \cite{Kurkela:2009gj} but fully proven only recently \cite{Ghisoiu:2016swa}, and provide a systematic and easily automatizable way of performing all $q_0$ integrations in the individual graphs (see also a related result presented in \cite{Bugrii:1995vn}). This procedure, which is summarized in fig.~5 and thoroughly explained in \cite{Ghisoiu:2016swa}, reduces the evaluation of a scalarized Euclidean Feynman diagram at finite chemical potential into a number of three-dimensional phase space integrals over on-shell amplitudes, where the entire $\mu$-dependence resides in the 3d integration measure.

\begin{figure}[t]
\centering
\includegraphics[width=12cm,clip]{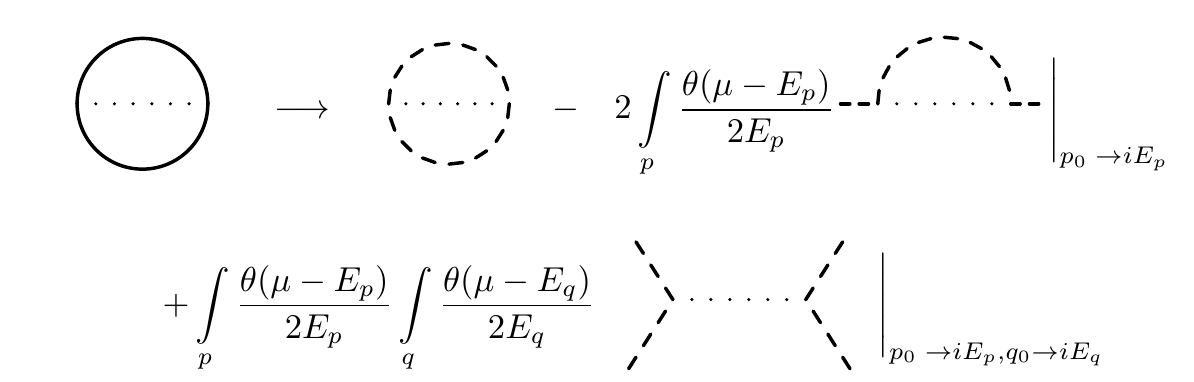}
\caption{An illustration of the cutting procedure of \cite{Ghisoiu:2016swa} being applied to a two-loop vacuum graph. Here, the solid line denotes a scalarized fermion propagator containing a chemical potential, the dotted line a boson propagator, and the dashed line a fermion propagator evaluated at $\mu=0$. We have abbreviated here $E_p\equiv \sqrt{p^2+m^2}$, with $m$ being the mass of the corresponding line, and note that the amplitudes are to be evaluated assuming the external momenta to be real-valued.}
\label{fig-5}       
\end{figure}

Using the cutting rules, the evaluation of the full set of four-loop Feynman diagrams in QCD is in practice divided to the following list of steps:
\begin{enumerate}
 \item Perform all Lorentz algebra and Dirac traces using programs of symbolic manipulation, such as FORM.
 \item Perform all $q_0$ momentum integrations with the help of the cutting rules.
 \item Use Integration by Parts and other well-established tools of perturbative quantum field theory to simplify the on-shell amplitudes produced by the application of the cutting rules. 
\item Evaluate the remaining amplitudes analytically and use a combination of analytical and numerical techniques to tackle the phase space integrations.
 \end{enumerate}
While the steps 1-3 here are in principle straightforward (albeit somewhat lengthy), the main challenge is clearly  the last step, which will likely require extensive efforts to be carried out.

\section{Conclusions}

In this conference proceedings contribution, we have briefly reviewed two recent results in the field of perturbative finite-density QCD \cite{Kurkela:2016was,Ghisoiu:2016swa}. The former generalized an existing three-loop EoS of zero-temperature quark matter to nonzero temperatures, while the latter provided a novel computational aid for perturbative calculations carried out at $T=0$ and $\mu\neq 0$. In the future, we plan to apply this computational tool, dubbed cutting rules, to the determination of the four-loop, or ${\mathcal O}(g^6)$, EoS of zero-temperature quark matter. In this context, we discussed the general outline of the required computations and in particular explained the procedure needed for accounting for the contributions of the IR sensitive soft degrees of freedom to the pressure.

Even if successful, obtaining the four-loop EoS of quark matter is  not guaranteed to dramatically improve the precision of the neutron star matter EoS. Firstly, it is nontrivial to quantitatively estimate in advance, how much obtaining the ${\mathcal O}(g^6)$ contribution will decrease the uncertainty of the quark matter EoS. Secondly, if the answer to this question is positive, we encounter the need to obtain a better estimate for the pairing contributions to the EoS, which become sizable below $\mu_\text{B}\approx 2$ GeV and are challenging to determine. And finally, even when equipped with an accurate quark matter EoS, dramatically shrinking the magnitude of the green region in fig.~1, where an interpolating EoS is needed, clearly requires advances also on the low-density side. Despite these potential problems, we nevertheless strongly feel that the importance of the challenge being faced --- obtaining an accurate theoretical understanding of the bulk properties of neutron star matter using only first principles machinery --- is enough to warrant extensive efforts on the pQCD front.


\begin{thebibliography}{}

\bibitem{Lattimer:2004pg}
  J.~M.~Lattimer and M.~Prakash,
  Science {\bf 304} (2004) 536
  [astro-ph/0405262].

  


\bibitem{Brambilla:2014jmp}
  N.~Brambilla {\it et al.},
  Eur.\ Phys.\ J.\ C {\bf 74} (2014) 10,  2981
  [arXiv:1404.3723 [hep-ph]].

\bibitem{deForcrand:2010ys}
  P.~de Forcrand,
  PoS LAT {\bf 2009} (2009) 010
  [arXiv:1005.0539 [hep-lat]].
  
  
\bibitem{Tews:2012fj}
  I.~Tews, T.~Kruger, K.~Hebeler and A.~Schwenk,
  Phys.\ Rev.\ Lett.\  {\bf 110} (2013) 3,  032504
  [arXiv:1206.0025 [nucl-th]].
  
  
\bibitem{Drews:2016wpi}
  M.~Drews and W.~Weise,
  arXiv:1610.07568 [nucl-th].
  
  
\bibitem{Buballa:2003qv}
  M.~Buballa,
  Phys.\ Rept.\  {\bf 407} (2005) 205
  [hep-ph/0402234].
  
\bibitem{Li:2015uea}
  S.~w.~Li, A.~Schmitt and Q.~Wang,
  Phys.\ Rev.\ D {\bf 92} (2015) no.2,  026006
  [arXiv:1505.04886 [hep-ph]].
  
\bibitem{Preis:2016fsp}
  F.~Preis and A.~Schmitt,
  JHEP {\bf 1607} (2016) 001
  [arXiv:1606.00675 [hep-ph]].
  
\bibitem{Hoyos:2016zke}
  C.~Hoyos, D.~Rodriguez Fernandez, N.~Jokela and A.~Vuorinen,
  Phys.\ Rev.\ Lett.\  {\bf 117} (2016) no.3,  032501
  [arXiv:1603.02943 [hep-ph]].
  
\bibitem{Hoyos:2016cob}
  C.~Hoyos, N.~Jokela, D.~Rodriguez Fernandez and A.~Vuorinen,
  arXiv:1609.03480 [hep-th].
  
\bibitem{Kurkela:2014vha}
  A.~Kurkela, E.~S.~Fraga, J.~Schaffner-Bielich and A.~Vuorinen,
  Astrophys.\ J.\  {\bf 789} (2014) 127
  [arXiv:1402.6618 [astro-ph.HE]].
  
  
  
  
\bibitem{annalanattila}  
  E.~Annala and J.~Nattila, In preparation.
  
\bibitem{Negele:1971vb}
  J.~W.~Negele and D.~Vautherin,
  Nucl.\ Phys.\ A {\bf 207} (1973) 298.
  
\bibitem{Ruester:2005fm}
  S.~B.~Ruester, M.~Hempel and J.~Schaffner-Bielich,
  Phys.\ Rev.\ C {\bf 73} (2006) 035804
  [astro-ph/0509325].

 
\bibitem{Kurkela:2009gj}
  A.~Kurkela, P.~Romatschke and A.~Vuorinen,
  Phys.\ Rev.\ D {\bf 81} (2010) 105021
  [arXiv:0912.1856 [hep-ph]].
 


  
   
\bibitem{Drischler:2016djf}
  C.~Drischler, A.~Carbone, K.~Hebeler and A.~Schwenk,
  Phys.\ Rev.\ C {\bf 94} (2016) no.5,  054307
  [arXiv:1608.05615 [nucl-th]].

  




\bibitem{Borsanyi:2010cj}
  S.~Borsanyi, G.~Endrodi, Z.~Fodor, A.~Jakovac, S.~D.~Katz, S.~Krieg, C.~Ratti and K.~K.~Szabo,
  JHEP {\bf 1011} (2010) 077
  [arXiv:1007.2580 [hep-lat]].
  
\bibitem{Borsanyi:2013bia}
  S.~Borsanyi, Z.~Fodor, C.~Hoelbling, S.~D.~Katz, S.~Krieg and K.~K.~Szabo,
  Phys.\ Lett.\ B {\bf 730} (2014) 99
  [arXiv:1309.5258 [hep-lat]].

  

\bibitem{klry}
K.~Kajantie, M.~Laine, K.~Rummukainen and Y.~Schr\"oder,
Phys.\ Rev.\ D {\bf 67} (2003) 105008 [hep-ph/0211321].



  
  
\bibitem{Fraga:2013qra}
  E.~S.~Fraga, A.~Kurkela and A.~Vuorinen,
  Astrophys.\ J.\  {\bf 781} (2014) 2,  L25
  [arXiv:1311.5154 [nucl-th]].


\bibitem{Andersen:2011sf}
  J.~O.~Andersen, L.~E.~Leganger, M.~Strickland and N.~Su,
  JHEP {\bf 1108} (2011) 053
  [arXiv:1103.2528 [hep-ph]].

  


  
  
  
\bibitem{fmcl}
B.~A.~Freedman and L.~D.~McLerran,
Phys.\ Rev.\ D {\bf 16} (1977) 1169;
  V.~Baluni,
  Phys.\ Rev.\ D {\bf 17} (1978) 2092.

\bibitem{avpres}
A.~Vuorinen,
Phys.\ Rev.\ D {\bf 68} (2003) 054017
[hep-ph/0305183].

 
\bibitem{Hebeler:2013nza}
  K.~Hebeler, J.~M.~Lattimer, C.~J.~Pethick and A.~Schwenk,
  Astrophys.\ J.\  {\bf 773} (2013) 11
  [arXiv:1303.4662 [astro-ph.SR]].
 


\bibitem{Shen:1998gq}
  H.~Shen, H.~Toki, K.~Oyamatsu and K.~Sumiyoshi,
  Nucl.\ Phys.\ A {\bf 637} (1998) 435
  [nucl-th/9805035].


  
  
\bibitem{Ipp:2006ij}
  A.~Ipp, K.~Kajantie, A.~Rebhan and A.~Vuorinen,
  Phys.\ Rev.\ D {\bf 74} (2006) 045016
  [hep-ph/0604060].

\bibitem{Kurkela:2016was}
  A.~Kurkela and A.~Vuorinen,
  Phys.\ Rev.\ Lett.\  {\bf 117} (2016) no.4,  042501
  [arXiv:1603.00750 [hep-ph]].

    
\bibitem{Laine:2011xm}
  M.~Laine, A.~Vuorinen and Y.~Zhu,
  JHEP {\bf 1109} (2011) 084
  [arXiv:1108.1259 [hep-ph]].


  
\bibitem{Appelquist:1981vg}
  T.~Appelquist and R.~D.~Pisarski,
  Phys.\ Rev.\ D {\bf 23} (1981) 2305.

\bibitem{Kajantie:1995dw}
  K.~Kajantie, M.~Laine, K.~Rummukainen and M.~E.~Shaposhnikov,
  Nucl.\ Phys.\ B {\bf 458} (1996) 90
  [hep-ph/9508379].

\bibitem{Braaten:1995cm}
  E.~Braaten and A.~Nieto,
  Phys.\ Rev.\ D {\bf 51} (1995) 6990
  [hep-ph/9501375].

\bibitem{Braaten:1989mz}
  E.~Braaten and R.~D.~Pisarski,
  Nucl.\ Phys.\ B {\bf 337} (1990) 569.
  
\bibitem{Braaten:1991gm}
  E.~Braaten and R.~D.~Pisarski,
  Phys.\ Rev.\ D {\bf 45} (1992) 1827.

  
    
\bibitem{Andersen:1999sf}
  J.~O.~Andersen, E.~Braaten and M.~Strickland,
  Phys.\ Rev.\ D {\bf 61} (2000) 014017
  [hep-ph/9905337].
  
  
\bibitem{Gerhold:2004tb}
  A.~Gerhold, A.~Ipp and A.~Rebhan,
  Phys.\ Rev.\ D {\bf 70} (2004) 105015
  [hep-ph/0406087].
  


  
\bibitem{Gerhold:2005uu}
  A.~Gerhold and A.~Rebhan,
  Phys.\ Rev.\ D {\bf 71}, 085010 (2005)
  [hep-ph/0501089].

  

\bibitem{newpaper}
  I.~Ghisoiu, T.~Gorda, A.~Kurkela, P.~Romatschke and A.~Vuorinen,
  In preparation.


 
\bibitem{Ghisoiu:2016swa}
  I.~Ghisoiu, T.~Gorda, A.~Kurkela, P.~Romatschke, M.~Sappi and A.~Vuorinen,
  arXiv:1609.04339 [hep-ph].
  
\bibitem{Bugrii:1995vn}
  A.~I.~Bugrii and V.~N.~Shadura,
  hep-th/9510232.
  



\end{thebibliography}
\end{document}